\documentclass[preprint,showpacs,preprintnumbers,amsmath,amssymb,prl]{revtex4}
\usepackage[pdftex]{graphicx}
\usepackage{bm}
\usepackage{float}
\usepackage{color}
\usepackage{hyperref}
\usepackage{pdfpages}
\usepackage[T1]{fontenc}
\begin{document}
\title{\color{blue}Investigation of the dynamical slowing down process in soft glassy colloidal suspensions: comparisons with supercooled liquids}
\author{Debasish Saha}
\email{debasish@rri.res.in}
\affiliation{Soft Condensed Matter Group, Raman Research Institute, C. V. Raman Avenue, Sadashivanagar, Bangalore 560 080, INDIA}
\author{Yogesh M Joshi}
\email{joshi@iitk.ac.in}
\affiliation{Department of Chemical Engineering, Indian Institute of Technology Kanpur, Kanpur 208 016, INDIA.}
\author{Ranjini Bandyopadhyay}
\email{ranjini@rri.res.in}
\affiliation{Soft Condensed Matter Group, Raman Research Institute, C. V. Raman Avenue, Sadashivanagar, Bangalore 560 080, INDIA}
\vspace{0.5cm}
\date{\today}
\begin{abstract}
The primary and secondary relaxation timescales of aging colloidal suspensions of Laponite are estimated from intensity autocorrelation functions obtained in dynamic light scattering (DLS) experiments. The dynamical slowing down of these relaxation processes are compared with observations in fragile supercooled liquids  by establishing a one-to-one mapping between the waiting time since filtration of a Laponite suspension and the inverse of the temperature of a supercooled liquid that is rapidly quenched towards its glass transition temperature. New timescales, such as the Vogel time and the Kauzmann time, are extracted to describe the phenomenon of dynamical arrest in Laponite suspensions. In results that are strongly reminiscent of those extracted from supercooled liquids approaching their glass transitions, it is demonstrated that the Vogel time calculated for each Laponite concentration is approximately equal to the Kauzmann time, and that a strong coupling exists between the primary and secondary relaxation processes of aging Laponite suspensions. Furthermore, the experimental data presented here clearly demonstrates the self-similar nature of the aging dynamics of Laponite suspensions within a range of sample concentrations.
\end{abstract}
\maketitle
\section{Introduction}
Relaxation processes in supercooled liquids are characterised by two-step decays \cite{gotze_mct}. The faster ($\beta$) decay corresponds to the rattling of the particle within a cage formed by its neighbours, while the slower ($\alpha$) decay corresponds to its cooperative diffusive dynamics between cages. The transport properties (i.e. viscosity, diffusivity etc.) and the relaxation timescales of a glass former change sharply as the glass transition is approached \cite{Ediger_Angell_Nagel_JPC_1996}. The primary or the $\alpha$-relaxation time becomes increasingly slow and diverges in the vicinity of the glass transition.  The dependence of this relaxation time on temperature in a strong glass former is nearly Arrhenius and the degree of deviation from Arrhenius behaviour is measured as `fragility'. For fragile glass formers, the $\alpha$-relaxation time shows a Vogel-Fulcher-Tammann ({\it VFT}) dependence on temperature ($T$), with the fragility index depending solely on the material \cite{Angell_Fragility_1991}. The density of potential energy minima of the configurational states in the potential energy landscape determines the strong or fragile behaviors of supercooled liquids \cite{Angell_JP_CS_1988}. Strong glasses have a lower density of minima and their entropy increases slowly with decreasing temperature, thereby resulting in nearly Arrhenius behavior \cite{Adam_Gibbs_JCP_1965}. On the contrary, fragile glasses have a larger density of minima which causes super-Arrhenius behavior of the $\alpha$-relaxation. Other secondary relaxation processes also simultaneously take place in the same temperature range. In supercooled liquids and molecular glasses, one of them is the Johari-Goldstein (JG) $\beta$-relaxation process \cite{Johari_Goldstein_JChemPhys_1970,Johari_Goldstein_JChemPhys_1971,Thayyil_Ngai_Philosophical_Magazine_2008,Ngai_JCP_1998}, which is the slowest of the secondary relaxation processes.\\
\indent In the last two decades, colloidal glasses have emerged as excellent model candidates for the study of glasses and glass formers. While supercooled liquids can be driven towards their glass transitions by rapidly quenching their temperatures, the glass transition in colloidal suspensions can be achieved by increasing the volume fraction $\phi$. For a colloidal suspension of hard spheres, increasing $\phi$ towards a glass transition volume fraction $\phi_{g}$ plays the same role as supercooling a liquid towards its glass transition temperature $T_{g}$ \cite{Pusey_Van_Megen_Nature_1986,Marshall_Zukoski_JPC_1990}.\\
\indent In recent years, colloidal glasses formed by the synthetic clay Laponite have been studied extensively \cite{bonn_epl,Ruzicka_2004,ruzicka_review,schosseler_PRE,kaloun_PRE,bandyopadhyay_prl,tanaka_pre,joshi_lang1,Negi_Osuji_PRE_2010,Angelini_SM_2013}. Interestingly, aging Laponite suspensions show many similarities with supercooled liquids and molecular glasses. These include the observation of well-separated fast and slow timescales \cite{Abou_Bonn_Meunier_PRE_2001}, the absence of thermorheological simplicity \cite{shahin_prl,lcstruik}, asymmetry in structural recovery following a step temperature change \cite{dhavale_SM,kovacs_jps}, probe size-dependent diffusion \cite{Strachan_et_al_PRE_2006} and the presence of more complex phenomena such as overaging \cite{lacks_prl,bandy_SM}. Laponite particles are monodisperse discs of diameter 25 nm and width 1 nm. In an aqueous medium, the dissociation of Na$^{+}$ ions from the Laponite platelet results in negatively charged faces, while the edge of the platelet acquires a charge that depends on the pH of the medium. At a pH of 10, the edge of the Laponite platelet is estimated to have a weak positive charge \cite{tawari_edge}. Overall, in an aqueous medium, Laponite particles interact {\it via} face-to-face, long range repulsions and edge-to-face, short range attractions \cite{tanaka_pre}. 
Remarkably, for $\phi >$ 0.004, their aqueous suspensions undergo ergodicity breaking over a duration of days, with the free-flowing liquid getting transformed into a soft solid phase that can support its own weight.\\
\indent Ruzicka {\it et al.} report the existence of two different concentration-dependent routes as Laponite clay suspensions approach the arrested state \cite{Ruzicka_2004}.  They claim that at high clay concentrations, the system forms a repulsive Wigner glass whose elementary units are single Laponite platelets, while at low clay concentrations (1.0 wt\% $< C_{w} <$ 2.0 wt\%), clusters of Laponite platelets form an attractive gel. Interestingly, recent work on this subject suggests that the influence of attractive interactions cannot be ruled out even at high Laponite concentrations \cite{ Ruzicka_PRL_2010,joshi_lang1}. Laponite suspensions also show very interesting phase behavior as the salt concentration is varied \cite{Ruzicka_Langmuir_2006,Ruzicka_Philosophical_Magazine_2007}. A gel or a glass state, and a nematic gel state are observed at low salt concentration as the clay concentration is increased. At very high ionic strengths, there is phase separation \cite{Tanaka_Bonn_PRE_2004}. Recent experimental observations and simulations in the gel state show that for very high waiting times, suspensions at weight concentration $C_{w}\leq$ 1.0 wt\% phase separate in the absence of salt into clay-rich and clay-poor phases, while suspensions at concentrations 1.0 wt\% $< C_{w} <$ 2.0 wt\% do not phase separate, giving rise to a true equilibrium gel obtained from an empty liquid \cite{Ruzicka_Nature_Materials_2011}.\\
\indent In an aging Laponite clay suspension, the effective volume fraction changes spontaneously and continuously with waiting time due to the spontaneous evolution of inter-particle electrostatic interactions \cite{bandyopadhyay_prl,bandyopadhyay_ssc,joshi_lang1}. In this work, we perform dynamic light scattering (DLS) experiments to measure the time-evolution of the primary and secondary relaxation processes of aging Laponite suspensions. We use our data to establish connections between aging Laponite suspensions undergoing dynamical arrest and fragile supercooled liquids approaching their glass transitions. We show here that increasing the waiting time $t_{w}$ of aging Laponite suspensions is equivalent to decreasing the thermodynamic temperature $T$ of supercooled liquids. While the Vogel-Fulcher-Tammann ({\it VFT}) functional form (with $1/T$ mapped with sample age $t_{w}$) was demonstrated to work for the slower $\alpha$-relaxation timescale of aging Laponite suspensions \cite{Ruzicka_2004}, we show here that $\beta$-relaxation follows an Arrhenius form (with $1/T$ also mapped with sample age $t_{w}$) as expected for supercooled liquids \cite{Ngai_JCP_1998,Stillinger_Science_1995}. A correspondence between temperature ($T$) and the waiting time since sample preparation ($t_{w}$) was reported in numerical studies of physical and chemical gelation \cite{Sciortino_et_al_Soft_Matter_2009} and in Monte Carlo simulations of  patchy-particle models of  Laponite discs \cite{Ruzicka_Nature_Materials_2011}. The role of thermodynamic temperature in the dynamical slowing down process of a colloidal glass produced by tethering polymers to the surface of inorganic nanoparticles has been investigated in the context of soft glassy rheology \cite{Archer_PRL_2011}.\\
\indent We next propose new timescales (the timescale $t^{\infty}_{\beta}$ associated with the fast process, the Vogel time $t^{\infty}_{\alpha}$ and the Kauzmann time $t_k$) to demonstrate several remarkable similarities that exist between supercooled liquids and soft glassy materials. We demonstrate a coupling between $t^{\infty}_{\beta}$ and the glass transition time $t_g$ \cite{Angell_Fragility_1991}. An analogous coupling between the glass transition temperature of a supercooled liquid and the activation energy corresponding to its $\beta$-relaxation process has been suggested and experimentally verified for supercooled liquids \cite{Kudlik_et_al_EPL_1997,Kudlik_et_al_J_Non_Cryst_Solids_1998, Vyazovkin_Dranca_Phermaceutical_Research_2006}, but has never been demonstrated in soft materials. We also show that a simple linear correlation exists between the Vogel time $t^{\infty}_{\alpha}$ and the Kauzmann time $t_k$. This result is strongly reminiscent of a previous observation in supercooled liquids, where the Kauzmann temperature $T_k$ has been shown to be approximately equal to the Vogel temperature $T_{0}$ {\cite{Jackle_Rep_Prog_Phys_1986}. We demonstrate the self-similar time-evolutions of the fast and slow relaxation times, the stretching exponent $\beta$, and the width and non-Gaussian parameter ($\alpha_{1}$ and $\alpha_{2}$) characterizing the distributions of the slow relaxation time with changing Laponite concentration. Finally, we show that the fragility index $D$ is concentration-independent and interpret this result in terms of the self-similar nature of the intricate potential energy landscape of aging Laponite suspensions.
\section{Materials, sample preparation and Experimental methods}
All the experiments reported in this work are performed with Laponite RD procured from Southern Clay Products. Before every experiment, Laponite powder is dried in an oven at $120^{\circ}$C for at least 16 hours. Appropriate amounts of powder are added slowly and carefully to double-distilled and deionized Millipore water of resistivity 18.2 M$\Omega$-cm. The mixture is stirred vigorously until it becomes optically clear. The resulting suspension is filtered using a syringe pump (Fusion 400, Chemyx Inc.) at a constant flow rate (3.0 ml/min) by passing through a 0.45 $\mu$m Millipore Millex-HV syringe-driven filter unit. The filtered suspension is loaded and sealed in a cuvette for DLS experiments. Laponite suspensions of concentrations 2.0\% w/v, 2.5\% w/v, 3.0\% w/v  and 3.5\% w/v are used in this study. Here, the concentration (\%w/v) is the weight of Laponite in 100 ml of water. The mechanical properties of all the suspensions evolve spontaneously with time and exhibit the typical signatures of soft glassy rheology \cite{miyazaki_epl}.\\
\indent The DLS experiments are performed with a Brookhaven Instruments Corporation (BIC) BI-200SM spectrometer equipped with a 150 mW solid state laser (NdYVO$_{4}$, Coherent Inc., Spectra Physics) having an emission wavelength of 532 nm. A refractive index-matching bath filled with decaline contains the cuvette filled with the sample. To avoid any kind of disturbance, the sample, once loaded,  is not removed until the end of the experiment. The temperature of the bath is maintained at $25^{\circ}$C by water circulation with a temperature controller (Polyscience Digital). A Brookhaven BI-9000AT Digital Autocorrelator is used to measure the intensity autocorrelation function of the light scattered from the samples. The intensity autocorrelation function $g^{(2)}(t)$ is defined as $g^{(2)}(t) = \frac{<I(0)I(t)>}{<I(0)>^{2}} =  1+ A|g^{(1)}(t)|^{2}$, \cite{bern_pecora}, where $I(t)$ is the intensity at a delay time $t$, $g^{(1)}(t)$ is the normalized electric field autocorrelation function, $A$ is the coherence  factor, and the angular brackets $< >$ represents an average over time. Experiments were performed at different scattering angles ($60^{\circ}$, $75^{\circ}$, $90^{\circ}$, $105^{\circ}$, $120^{\circ}$ and $135^{\circ}$). Data acquired at $90^{\circ}$ are reported in the manuscript. Some representative data acquired at $60^{\circ}$ is shown in supporting information. The duration of data collection is kept long enough (2-3 minutes) to ensure a large photon count ($>10^{7}$ counts/run). Details of the data analysis protocols used in this work, for example in the calculations of the width and non-Gaussian parameters ($\alpha_{1}$ and $\alpha_{2}$) characterizing the distributions of the slow relaxation times, have been provided in supporting information.
\section{Results and Discussions}
\begin{figure}
\centering
  \includegraphics[height=6cm]{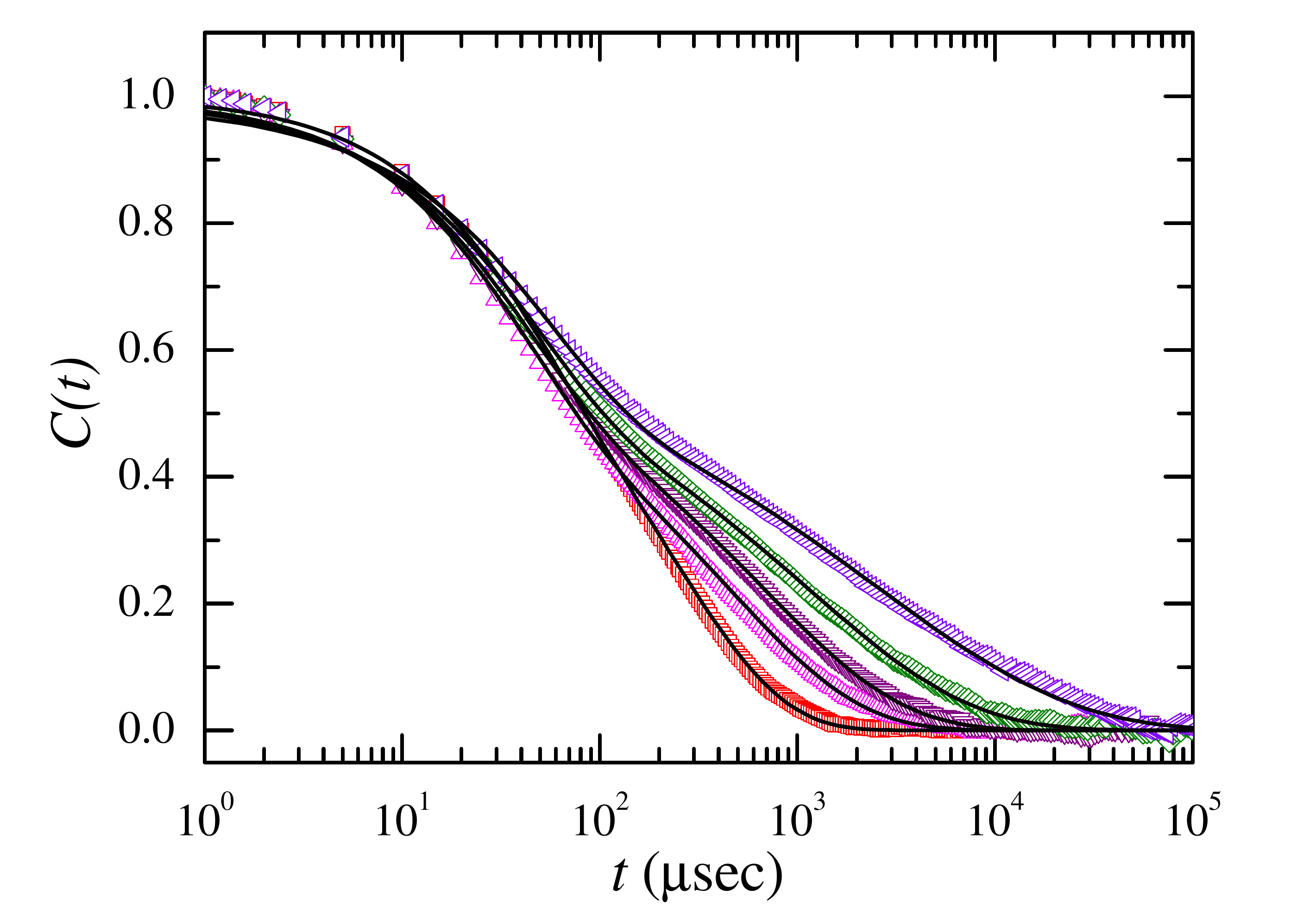}
  \caption{The normalized intensity autocorrelation functions $C(t)$, {\it vs.} the delay time $t$, at 25$^\circ$C and        scattering angle $\theta$ = 90$^\circ$ for 3.0\% w/v Laponite suspension at several different waiting times $t_{w}$ (from left to right): 0.5 hours ($\Box$), 6.0 hours ($\triangle$), 9.0 hours ($\nabla$), 12.0 hours ($\diamond$) and 15.0 hours ($   \triangleleft$). The solid lines are fits to equation 1.}
  \label{Figure 1}
\end{figure}
The relaxation dynamics of a medium can be analyzed by monitoring the temporal behavior of the intensity autocorrelation function $g^{(2)}(t)$. In figure 1, we plot the normalized intensity autocorrelation function, $C(t)=g^{(2)}(t)-1$, for a 3.0\% w/v Laponite suspension as a function of delay time, $t$, for experiments carried out at different waiting times $t_{w}$ since filtration of the sample. $C(t)$ shows a two-step decay, suggesting the presence of two distinct relaxation timescales. In addition, the decay in the autocorrelation function slows down progressively as the sample ages. For a glassy suspension, the two-step decay of $C(t)$ can be described as a squared sum of an exponential and a stretched exponential decay given by \cite{Ruzicka_2004}:
\begin{equation}
\label{equation 1}
C(t)=\left[a\exp\left\{-t/\tau_{1}\right\}+(1-a)\exp\left\{-(t/\tau_{ww})^{\beta}\right\}\right]^{2}
\end{equation}\
The fits to equation 1 (shown by the solid lines in figure 1) describe the decays of the normalized autocorrelation functions for a range of waiting times $t_{w}$ and for all the aging Laponite suspensions studied in this work. The fits are used to estimate the two relaxation timescales: $\tau_{1}$, the fast relaxation timescale that is associated with the secondary relaxation process, and $\tau_{ww}$, the slow timescale that is associated with the primary $\alpha$-relaxation process. In addition, the fits also yield values of the `stretching exponent', $\beta$, which is connected to the distribution of the $\alpha$-relaxation timescales.\\
\begin{figure}[!t]
\centering
	\includegraphics[width=8cm]{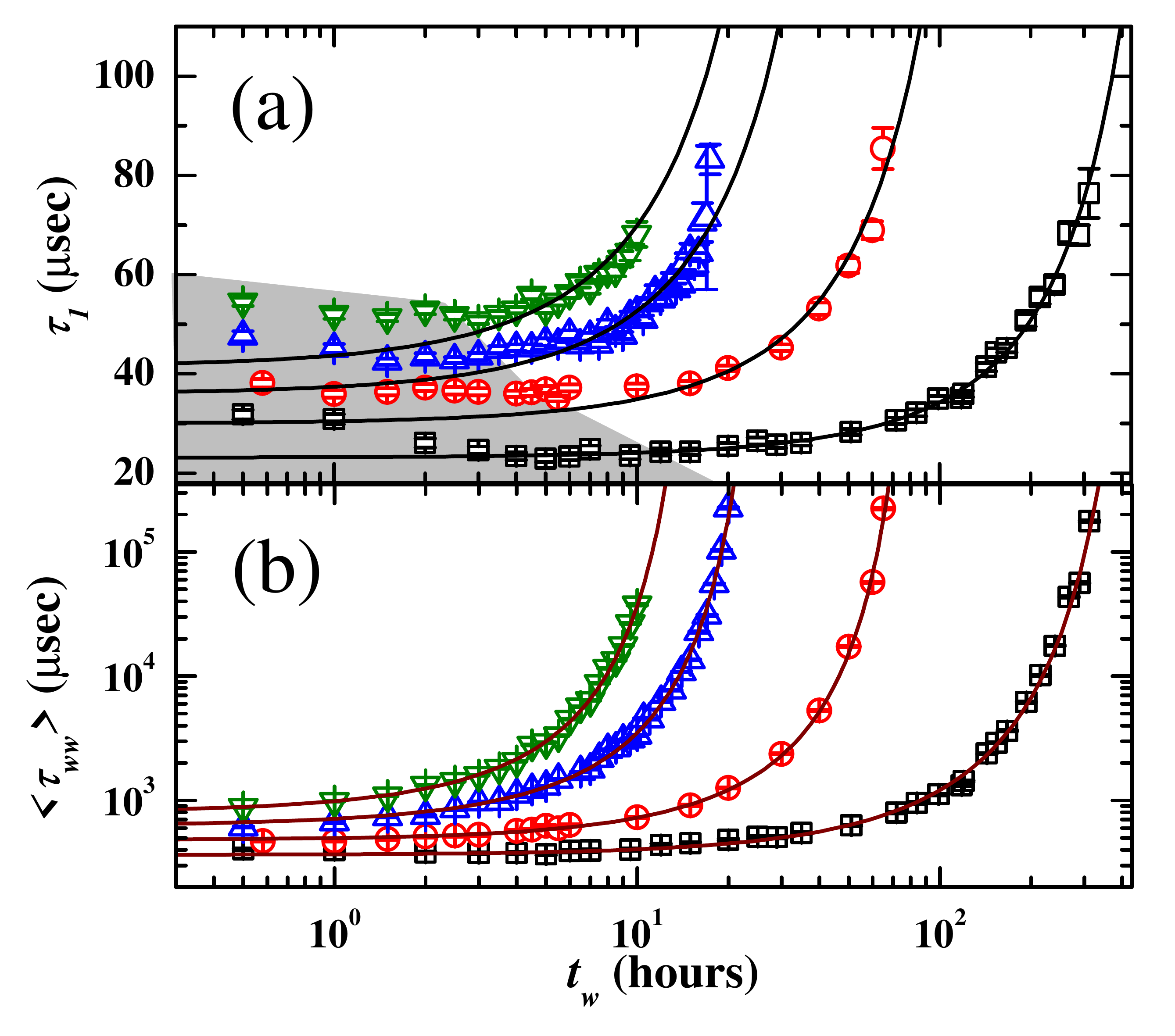}
	\caption{(a) The fast relaxation times, $\tau_{1}$, {\it vs.} waiting time, $t_{w}$, for Laponite samples prepared  at 25$^{\circ}$C and at concentration 2.0\% w/v ($\Box$), 2.5\% w/v ($\circ$), 3.0\% w/v ($\triangle$) and 3.5\% w/v  ($\nabla$). The solid lines show  fits to the modified Arrhenius functions, $\tau_{1}=\tau^{0}_{1}\exp(t_{w}/t^{\infty}_{\beta})$ (equation 2). Data are shifted vertically for better representation. The shaded portion highlights the initial decrease in $\tau_{1}$. (b) The mean $\alpha$-relaxation times, $<\tau_{ww}>$, {\it vs.} waiting time, $t_{w}$ are plotted for the same samples. The solid lines show fits to the modified {\it VFT} functions, $<\tau_{ww}>=<\tau_{ww}>^{0}\exp(Dt_{w}/(t^{\infty}_{\alpha}-t_{w}))$ (equation 3).}
	\label{Figure 2}
\end{figure}
\indent In figure 2(a), we plot the evolutions of $\tau_{1}$ with increasing $t_{w}$ for Laponite suspensions of different concentrations. Interestingly, $\tau_{1}$ evolves in two steps. At very small $t_{w}$, $\tau_{1}$ initially decreases (shown by the shaded portion in figure 2(a)), before increasing rapidly at large $t_{w}$. In addition, the evolution of $\tau_{1}$ shifts to smaller $t_{w}$ with increasing Laponite concentration.\\
\begin{figure}[!t]
\centering
	\includegraphics[width=3.5in]{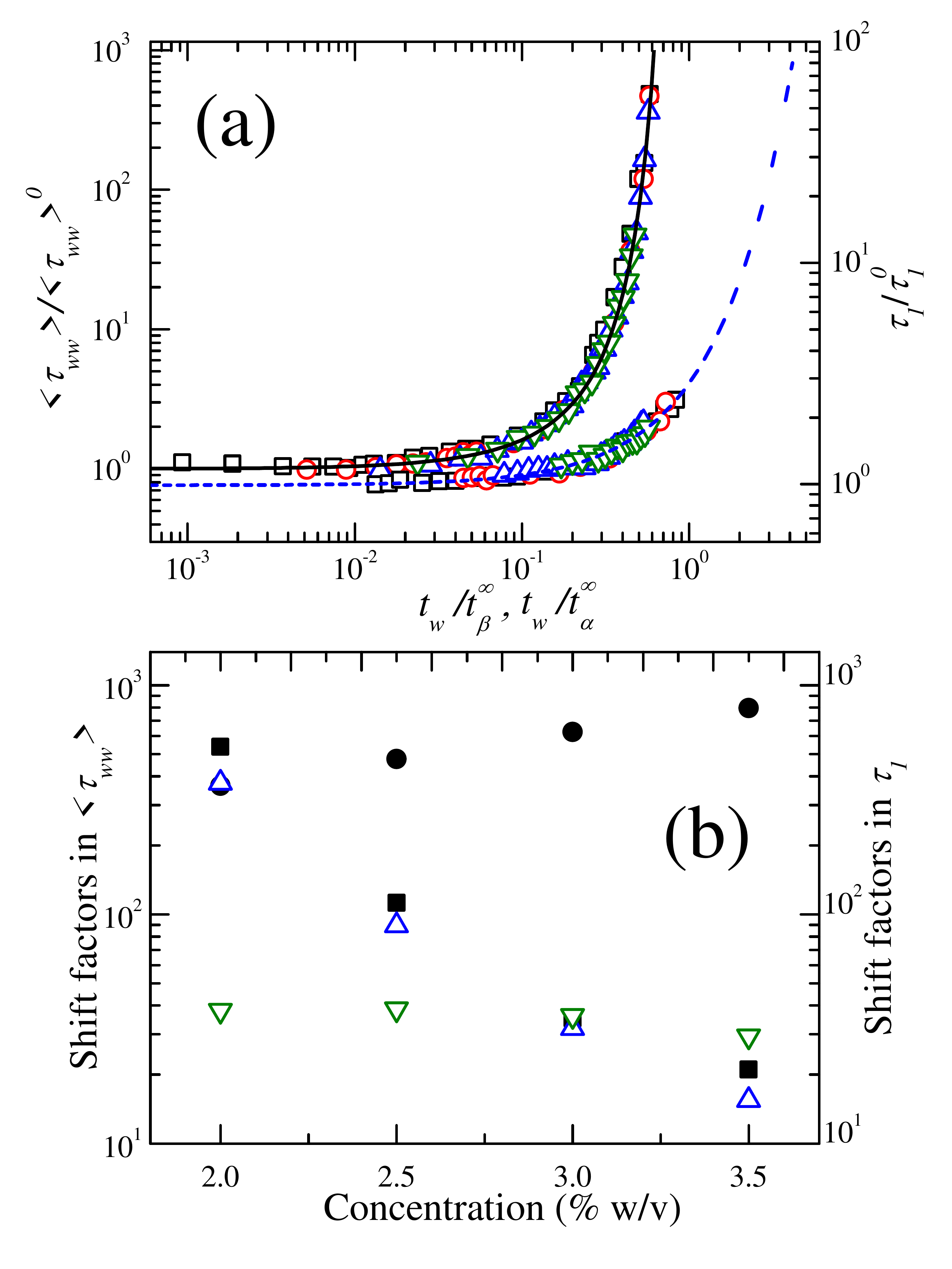}
	\caption{In (a), Superpositions of normalized $\tau_{1}$ and normalized $<\tau_{ww}>$  when plotted {\it vs.} $t_{w}/t^{\infty}_{\beta}$ and $t_{w}/t^{\infty}_{\alpha}$, respectively, for 2.0\% w/v ($\Box$), 2.5\% w/v ($\circ$), 3.0\% w/v ($\triangle$) and 3.5\% w/v ($\nabla$) Laponite suspensions. Dashed and solid lines are fits of normalized $\tau_{1}$ and normalized $<\tau_{ww}>$ to the modified Arrhenius and modified {\it VFT} functions (equations 2 and 3) respectively. In (b), the shift factors are plotted {\it vs.} Laponite concentration. The horizontal shift factors $t^{\infty}_{\beta}$ (hours) and $t^{\infty}_{\alpha}$ (hours), corresponding to the fast and slow relaxation processes respectively, are denoted by $\triangle$ and $\blacksquare$, respectively. The vertical shift factors, $\tau_{1}^{0}$ ($\mu$sec) and $<\tau_{ww}>^{0}$ ($\mu$sec), are denoted by $\nabla$ and $\bullet$, respectively.}
	\label{Figure 3}
\end{figure}
%
\begin{figure}[!t]
\includegraphics[width=3.5in]{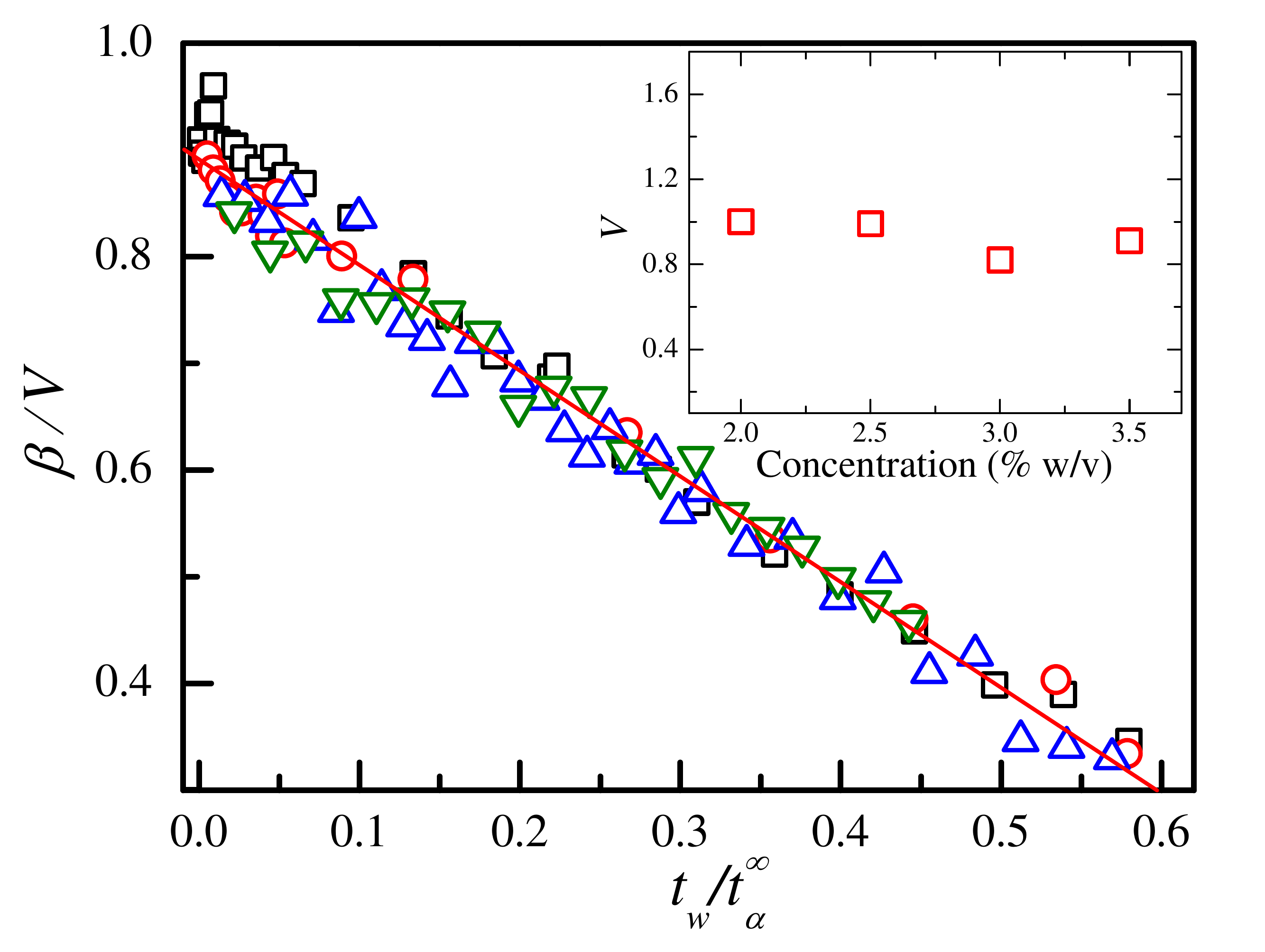}
\caption{Superposition of the normalized stretching coefficients $\beta$ when plotted {\it vs.} $t_{w}/t^{\infty}_{\alpha}$ for 2.0\% w/v ($\Box$), 2.5\% w/v ($\circ$), 3.0\% w/v ($\triangle$) and 3.5\% w/v ($\nabla$) Laponite suspensions. The straight line is a linear fit. In the inset: vertical shift factor ($V$) {\it vs.} Laponite concentration.}
\label{Figure 4}
\end{figure}
\indent Soon after mixing dry Laponite powder in water, hydration of clay takes place and water molecules diffuse into the interlayer gallery causing  the clusters to swell. Filtration of these suspensions breaks the clusters. After filtration, these broken clusters undergo further fragmentation  \cite{Joshi_JCP_2007}. In both cases, $\tau_{1}$ is expected to decrease until the swelling clusters or the fragmented parts undergo dynamical arrest due to strong inter-platelet interactions that evolve spontaneously \cite{Samim_Ranjini_Langmuir_2013}. The waiting time at which $\tau_{1}$ shows a minimum can therefore be considered as a measure of the time required for the onset of jamming. The waiting time associated with the minimum, $t_{w,min}$, decreases with increase in Laponite concentration (figure S1 in supporting information). As the Laponite concentration increases, the increase in the number of cage-forming particles can be associated with a decrease in the free space that is required for cage expansion and swelling of the  clusters. The minimum in $\tau_{1}$ ($t_{w,min}$) therefore shifts to smaller $t_{w}$ with increase in Laponite concentration.\\
\indent The slow timescale $\tau_{ww}$ is identified with the $\alpha$-relaxation process. The average value of $\tau_{ww}$, $<\tau_{ww}>  = ({\tau_{ww}}/{\beta})\Gamma({1}/{\beta})$, where $\Gamma$ is the Euler Gamma function \cite{Lindsey_JCP_1980}. In figure 2(b), the evolution of $<\tau_{ww}>$ is plotted as a function of $t_{w}$ for different concentrations of Laponite. In contrast to the non-monotonic behavior of $\tau_{1}$, $<\tau_{ww}>$ remains almost constant at small $t_{w}$. At larger $t_{w}$, $<\tau_{ww}>$ shows a sharp increase. Furthermore, the evolution of $<\tau_{ww}>$ shifts to larger $t_{w}$ with decrease in concentration of Laponite. The stretching exponents $\beta$ associated with $<\tau_{ww}>$ are obtained from fits of the data to equation 1 and are plotted as a function of $t_{w}$ in  figure S2 of supporting information. For small values of $t_{w}$, $\beta$ is close to unity. However with increase in $t_{w}$, $\beta$ decreases linearly, which signifies the broadening of the distribution associated with $<\tau_{ww}>$. The decrease in $\beta$ also shifts to smaller $t_{w}$ with increase in the concentration of Laponite.\\
\indent Because of self-similar curvatures in the evolutions of both $\tau_{1}$ (the monotonically increasing parts) and $<\tau_{ww}>$, the data plotted in figure 2 can be superposed upon horizontal and vertical shifting. This is shown in figure 3(a). The corresponding shift factors (the horizontal shift factors for $\tau_{1}$ and $<\tau_{ww}>$ are denoted by $t^{\infty}_{\beta}$ and $t^{\infty}_{\alpha}$ respectively, and the vertical shift factors for $\tau_{1}$ and $<\tau_{ww}>$ are denoted by $\tau^{0}_{1}$ and $<\tau_{ww}>^{0}$ respectively) are plotted in figure 3(b). It is observed that $<\tau_{ww}>^{0}$ ($\bullet$ in figure 3(b)) increases with Laponite concentration. This observation can be explained by considering that at higher concentrations, the particles are more easily confined in deep wells and can therefore be kinetically constrained at earlier times. This confirms that the sluggishness of the $\alpha$-relaxation process increases with increasing Laponite concentration. In addition to $\tau_{1}$ and $\tau_{ww}$, $\beta$ also shows superposition after appropriate shifting through a vertical shift factor ($V$) obtained as the value of $\beta$ at $t_{w}/t^{\infty}_{\alpha}\rightarrow0$. This is shown in figure 4.\\ 
\indent The self-similarity and sharp enhancement of $\tau_{1}$ and $<\tau_{ww}>$ with increase in $t_{w}$ are reminiscent of the changes that are observed in the fast ($\beta$) and slow ($\alpha$) timescales of supercooled liquids that are quenched rapidly towards their glass transition temperatures $T_{g}$ \cite{gotze_mct,Stillinger_Science_1995}. In supercooled liquids, the fast relaxation shows an Arrhenius dependence on temperature $T$ given by: $\tau_{1}=\tau^{0}_{1}\exp(E/k_{B}T)$. Here, $\tau^{0}_{1}$ is the fast relaxation time when $T\rightarrow\infty$, $E$ is the depth of the energy well associated with particle motion within the cage and $k_{B}$ is the Boltzmann constant. The slow $\alpha$-relaxation time, which represents the timescale associated with cage diffusion in supercooled liquids, demonstrates a dependence on temperature $T$ that is given by the Vogel-Fulcher-Tammann ({\it VFT}) law: $<\tau_{ww}> = <\tau_{ww}>^{0}\exp(DT_{0}/(T-T_{0}))$. Here, the temperature $T_{0}$ at which $<\tau_{ww}>$ diverges is called the Vogel temperature and $D$ is the fragility of the material. The Arrhenius equation is, therefore, a special case of the {\it VFT} equation in the limit $T_{0} \rightarrow$ 0 \cite{larson_rheology}. Clearly, for nonzero values of $T_{0}$, the slow timescale $<\tau_{ww}>$ diverges more rapidly than the fast timescale $\tau_{1}$.  In figure 3(a), we see a very similar situation, wherein $<\tau_{ww}>$ diverges much more rapidly when compared to $\tau_{1}$. It can therefore be appreciated that the slowdown observed in aqueous Laponite suspensions is equivalent to that seen in supercooled liquids, with the inverse of the temperature ($1/T$) in the latter case mapped with the waiting time ($t_{w}$) in the former. In order to assess the validity of the proposed mapping, we write a modified Arrhenius equation:
\begin{equation}
\label{eq:2}
\tau_{1}=\tau^{0}_{1}\exp(t_{w}/t^{\infty}_{\beta})
\end{equation}
Here, $t^{\infty}_{\beta}$ is a characteristic timescale associated with the slowdown of the fast relaxation process. Similarly, the modified {\it VFT} equation for the mean $\alpha$-relaxation time is written as:
\begin{equation}
\label{eq:3}
<\tau_{ww}> = <\tau_{ww}>^{0}\exp(Dt_{w}/(t^{\infty}_{\alpha}-t_{w})),
\end{equation}
where $t^{\infty}_{\alpha}$ is identified as a Vogel time and $<\tau_{ww}>$ is calculated from the distribution of slow relaxation times $\rho_{ww}(\tau)$ which is obtained by inverting the stretched exponential part of the autocorrelation decay. The expression for $\rho_{ww}(\tau)$ is given by \cite{Lindsey_JCP_1980},
\begin{equation}
\rho_{ww}(\tau)=-\frac{\tau_{ww}}{\pi\tau^2}\sum^{\infty}_{k=0}\frac{(-1)^k}{k!}\sin(\pi\beta k)\Gamma(\beta k+1)\left(\frac{\tau}{\tau_{ww}}\right)^{\beta k+1}
\label{equation 2}
\end{equation}
\begin{figure*}[!t]
\includegraphics[width=7in]{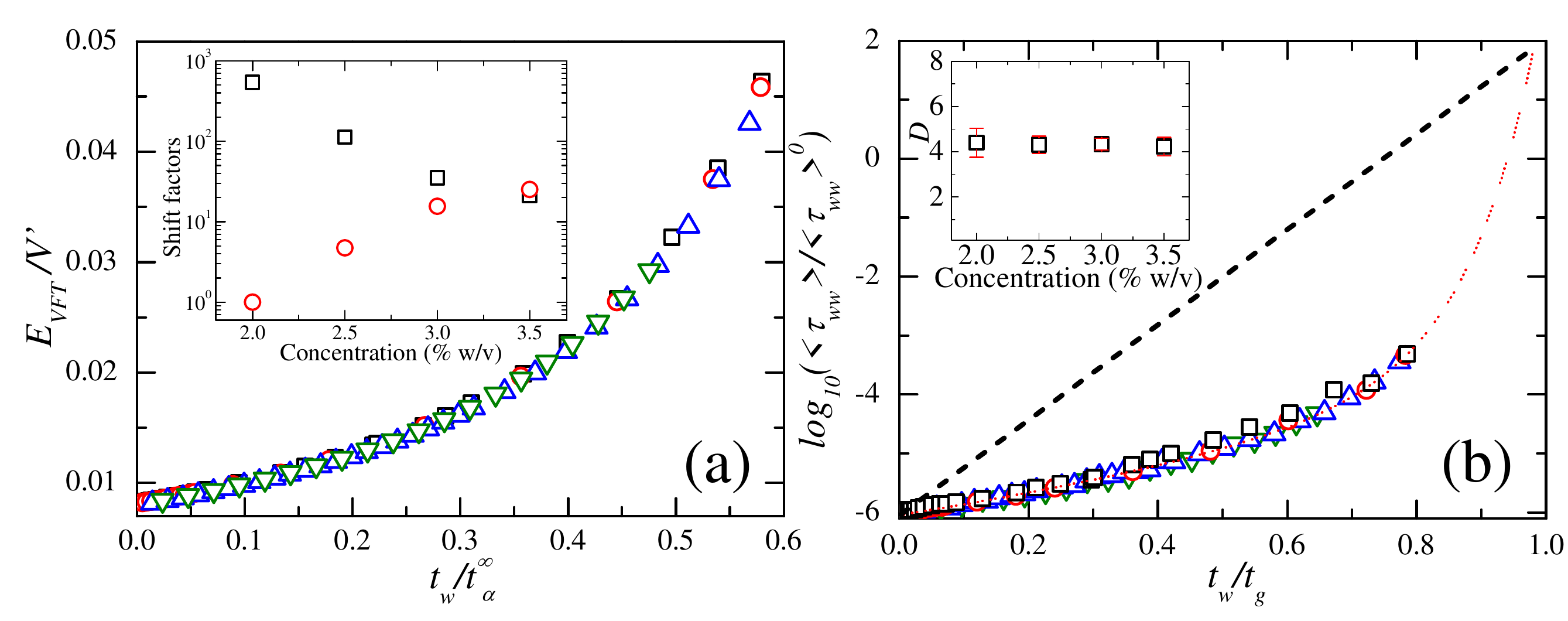}
\caption{(a). Superposition of the normalized apparent activation energies ($E_{VFT}$) associated with the $\alpha$-relaxation processes as a function of  ($t_{w}/t^{\infty}_{\alpha}$) for 2.0\% w/v ($\Box$), 2.5\% w/v ($\circ$), 3.0\% w/v ($\triangle$) and 3.5\% w/v ($\nabla$) Laponite suspensions. Inset shows the horizontal ($t^{\infty}_{\alpha}$ denoted by $\Box$) and vertical   ($V^{\prime}$ denoted by $\circ$) shift factors {\it vs.} Laponite concentration. (b) Angell plot for 2.0\% w/v ($\Box$), 2.5\% w/v ($\circ$), 3.0\% w/v ($\triangle$) and 3.5\% w/v ($\nabla$) Laponite suspensions. The dashed diagonal straight line is the Angell plot for strong supercooled liquids, while the dotted curve is for a fragile glassformers. In the inset, fragility index ($D$) is plotted {\it vs.} concentration of Laponite suspensions.}
\label{Figure 5}
\end{figure*}
\begin{figure*}[!t]
\includegraphics[width=7in]{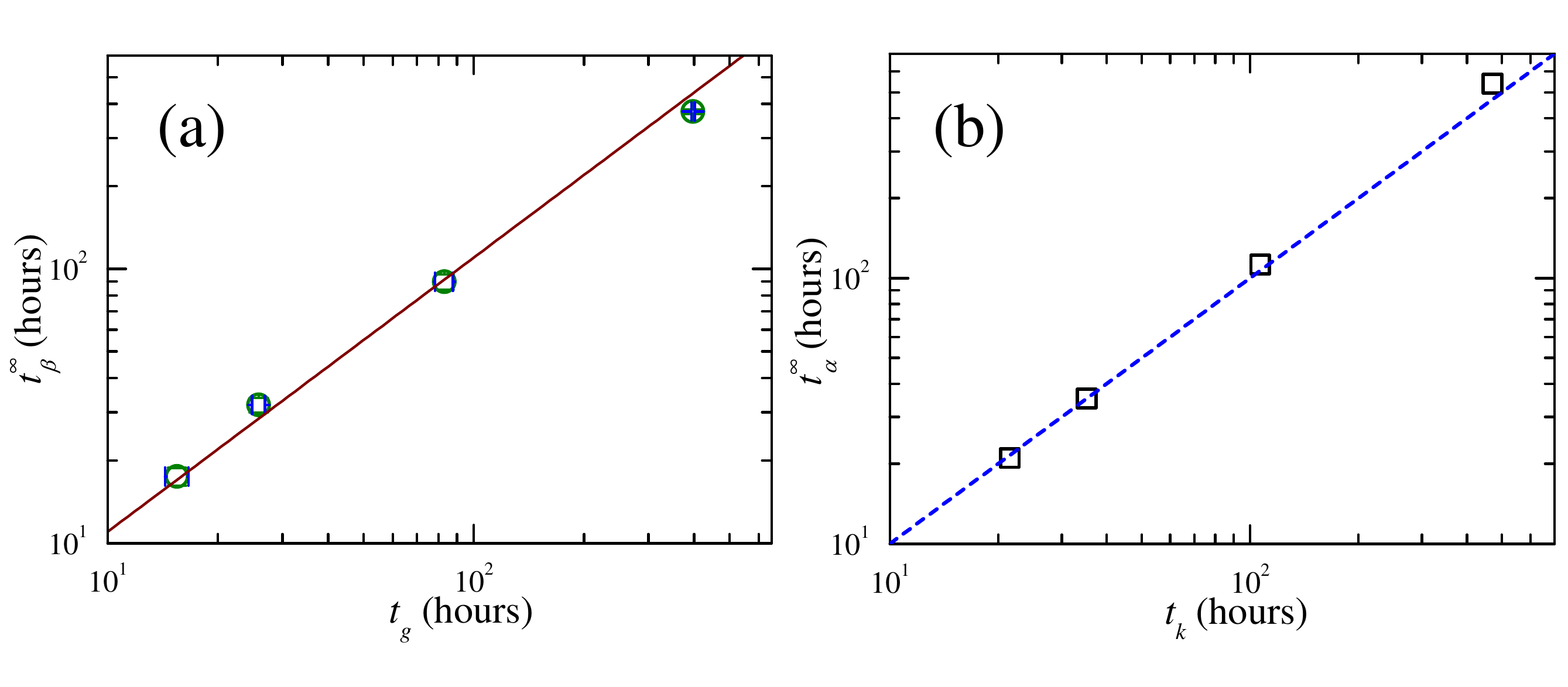}
\caption{(a) The fast relaxation timescale $t^{\infty}_{\beta}$ {\it vs.} the glass transition time $t_{g}$ (from left to right - 3.5\% w/v, 3.0\% w/v, 2.5\% w/v and 2.0\% w/v). The solid line ($t^{\infty}_{\beta} \approx$ (1.10 $\pm$ 0.05)$t_{g}$)  is a linear fit passing through origin. (b) The Vogel time, $t^{\infty}_{\alpha}$, is plotted {\it vs.} the Kauzmann time, $t_{k}$ (from left to right - 3.5\% w/v, 3.0\% w/v, 2.5\% w/v and 2.0\% w/v). The dashed line is a linear fit ($t^{\infty}_{\alpha} \approx t_{k}$)  passing through origin.}
\label{Figure 6}
\end{figure*}
In equations 2 and 3, the inverse of temperature $1/T$ in the Arrhenius and the {\it VFT} forms for supercooled liquids is mapped with $t_{w}$ and $1/T_{0}$ is mapped with $t^{\infty}_{\alpha}$. It can be seen in figures 2 and 3 that equations 2 and 3  fit the time-evolution of the  $\tau_{1}$ and $<\tau_{ww}>$ data extremely well.\\ 
\indent The strong dependence of $t^{\infty}_{\alpha}$ on Laponite concentration ($\blacksquare$ in figure 3(b)) can be explained by using a purely entropic picture. For $N$  particles, the configurational entropy $S_{c}\equiv Nk_{B}\ln\Omega(N)$, where $\Omega$ is the number of minima in the potential energy surface and $k_{B}$ is the Boltzmann constant, while the total number of configurational states is $\propto N!\exp(\alpha N)$, where $\alpha$ is a positive number \cite{Stillinger_Science_1995,Sastry_Nature_2001}. As $N$ increases with Laponite concentration, the total number of configurational states and the configurational entropy  $S_{c}$  increase rapidly with $N$.  This is accompanied by a decrease in the excluded volume available to the system. Increasing Laponite concentration therefore results in a decrease in $t^{\infty}_{\alpha}$, with the system being driven towards the glass transition at smaller waiting times.  In contrast, $\tau^{0}_{1}$ ($\nabla$ in figure 3(b)) remains approximately independent of concentration as the local environment of a particle trapped in a cage does not change appreciably with change in concentration.\\
\indent In an activated process, the dependence of a characteristic timescale on temperature is used to calculate the activation energy associated with that relaxation phenomenon. For an Arrhenius relaxation process represented by $\tau_{1}=\tau^{0}_{1}\exp(E/k_{B}T)$, $E$ is the activation energy and $k_{B}$ is the Boltzmann constant. For a {\it VFT} relaxation process described by $<\tau_{ww}>=<\tau_{ww}>^{0}\exp(DT_{0}/(T-T_{0}))$, the apparent activation energy is given by: $E_{VFT}=k_{B}DT_{0}T^{2}/(T-T_{0})^2$ \cite{Ediger_Angell_Nagel_JPC_1996,Introduction_to_Polymer_Viscoelasticity}. The activation energies associated with the modified Arrhenius and {\it VFT} processes in aging Laponite suspensions can be estimated by comparing with the corresponding relations for a supercooled liquid, with $1/T$ mapped with $t_{w}$ and $1/T_{0}$ with $t^{\infty}_{\alpha}$. These calculations, the details of which are supplied in supporting information, yield the following results:
\begin{equation}
E=(k_{B} c_{1})/t^{\infty}_{\beta}
\end{equation}
and
\begin{equation}
E_{VFT}=(k_{B}c_{2})[D t^{\infty}_{\alpha}/(t^{\infty}_{\alpha}-t_{w})^{2}]
\end{equation}
\indent Here, equation 5 represents activation energy ($E$) associated with $\tau_{1}$, while equation 6 represents the apparent activation energy ($E_{VFT}$) associated with $<\tau_{ww}>$. In these equations, $k_{B}$ is the Boltzmann constant, $D$ is the fragility parameter, and $c_{1}$ and $c_{2}$ are constants with dimensions [time]$\times$[temperature]. It can be seen in the inset of figure S3 of supporting information that the activation energy $E$ associated with $\tau_{1}$ is independent of $t_{w}$ and shows a power-law dependence on concentration of Laponite $c$ ($E\propto c^{5.7\pm0.3}$). $E_{VFT}$, associated with $<\tau_{ww}>$, on the other hand, remains constant at small $t_{w} (<<t^{\infty}_{\alpha}$), but shows a strong dependence on $t_{w}$ for large $t_{w}$ (figure S3 of supporting information). In addition, $E_{VFT}$ shifts to smaller waiting times with increase in concentration of Laponite. This agrees with our earlier results that Laponite suspensions of higher concentrations are driven faster towards an arrested state. Furthermore, our data implies that the evolution of the potential energy landscape with increasing $t_{w}$ is governed only by the $\alpha$-relaxation process. The self-similar nature of $E_{VFT}$ with changes in Laponite concentration  is apparent when the data is scaled appropriately (figure 5(a)). The same horizontal shift factor $t^{\infty}_{\alpha}$, used earlier to superpose the $<\tau_{ww}>$ data, is also used here.\\
\indent Following the definition proposed by Angell for supercooled liquids, we define the glass transition time $t_{g}$ as the time since sample preparation at which $<\tau_{ww}>$ = $100$ seconds for each Laponite concentration \cite{Angell_Fragility_1991}. The Angell plot corresponding to the $\alpha$-process of Laponite suspensions is shown in figure 5(b). Our data shows the same behaviour expected for fragile supercooled liquids (shown by the dotted line, where $1/T$ is mapped with $t_{w}$ as discussed in equations 2 and 3). The straight dashed line corresponds to strong glassformers for which the $\alpha$-relaxation timescale shows Arrhenius behavior. We identify the constant parameter $D$ in equation 3 as the fragility index \cite{Angell_Fragility_1991,Debenedetti_Nature_2001,Sastry_Nature_2001}. It is observed that the value of $D$ remains almost constant over the Laponite concentration range explored here (inset of figure 5(b)).  It has been pointed out that caged particles can get trapped in deeper energy wells with increase in the concentration of a glassformer \cite{Goldstein_JCP_1969}. However, our observation that $D$ is independent of Laponite concentration suggests that the overall topology of the potential energy landscape of aging Laponite suspensions remains unchanged even when Laponite concentration is changed \cite{Stillinger_Science_1995}.\\
\indent The simultaneous enhancements of the fast and slow timescales at high $t_{w}$ suggests the possibility that both these processes are strongly correlated with each other. In figure 6(a), the timescale $t^{\infty}_{\beta}$ associated with the fast relaxation process and obtained from fits to equation 2 is plotted {\it vs.} the glass transition time $t_{g}$. It is observed that these two timescales are strongly coupled. A linear fit to the data (solid line in figure 6(a)) yields $t^{\infty}_{\beta}=(1.10\pm0.05)t_{g}$. For supercooled liquids, the activation energy associated with the $\beta$-relaxation process was demonstrated to be proportional to the glass transition temperature $T_{g}$, with the exact relationship being given by $E_{\beta}=(24\pm3)RT_{g}$, where $R$ is the universal gas constant \cite{Kudlik_et_al_EPL_1997,Kudlik_et_al_J_Non_Cryst_Solids_1998,Vyazovkin_Dranca_Phermaceutical_Research_2006}. The relation obtained here between $t^{\infty}_{\beta}$ and $t_{g}$  is therefore strikingly similar to the observation in supercooled liquids. The fast relaxation process in Laponite glasses has previously been identified as a $\beta$-relaxation process \cite{Abou_Bonn_Meunier_PRE_2001}. The coupling between $t^{\infty}_{\beta}$ and $t_{g}$ seen in figure 6(a) is reminiscent of the behaviour seen in supercooled liquids, where the Johari-Goldstein (JG) $\beta$-relaxation  is seen to be coupled with the  $\alpha$-relaxation \cite{Ngai_JCP_1998,Kudlik_et_al_EPL_1997,Kudlik_et_al_J_Non_Cryst_Solids_1998,Vyazovkin_Dranca_Phermaceutical_Research_2006}. The assymetric nature of the Laponite particles, the observed Arrhenius dependence of the fast relaxation timescale on $t_{w}$ (figures 2(a) and 3(a)), and the decrease of the stretching exponent $\beta$ with increasing $\log(<\tau_{ww}>/\tau_{1})$ (figure S4 of supporting information), indicate a qualitative similarity of the fast relaxation process observed here and the Johari-Goldstein (JG) $\beta$-relaxation process reported in supercooled liquids \cite{Ngai_JCP_1998}. We therefore  speculate that the fast relaxation process observed in our experiments could be a JG $\beta$-relaxation process.\\
\indent The linear decrease of $\beta$ with $t_{w}$ (figure 4 and figure S2 in supporting information) is similar to the observation in fragile supercooled liquids where $\beta$  decreases linearly with $1/T$ \cite{Dixon_Nagel_PRL_1998}. We define a Kauzmann time $t_{k}$, as an analog of the Kauzmann temperature $T_{k}$ for supercooled liquids, by extrapolating the waiting time $t_{w}$ to the value at which $\beta\rightarrow$0 \cite{Bohmer_Ngai_Angell_JCP_1993}. The linear correlation that is obtained between $t^{\infty}_{\alpha}$ and $t_{k}$ ($t^{\infty}_{\alpha}\approx t_{k}$) is plotted in figure 6(b). This is strongly reminiscent of the behaviour of supercooled liquids where an analogous relationship ($T_{0}\approx T_{k}$) holds \cite{Jackle_Rep_Prog_Phys_1986}. We have tabulated the values of $t^{\infty}_{\alpha}$, $t_{k}$, $t^{\infty}_{\beta}$ and $t_{g}$, estimated for the four different Laponite concentrations, in table T1 of supporting information.\\
\indent We now analyze the distributions of the $\alpha$-relaxation timescales for various Laponite concentrations.  The distributions of the $\alpha$-relaxation timescales for a 3.0\% w/v Laponite suspension, $\rho_{ww}(\tau)$, at four different $t_{w}$ values (2 hr, 5 hr, 10 hr and 20 hr) are estimated using equation 4 and are plotted in the inset of figure 7(a). In all the samples studied, the distributions broaden significantly with increasing waiting time $t_{w}$.  We define a width parameter $\alpha_{1}$ as a measure of the broadening of $\rho_{ww}(\tau)$. The values of $\alpha_{1}$ calculated by us (details of the calculation of $\alpha_{1}$ and table T2 of our estimates of $\alpha_{1}$ values are supplied in supporting information) are seen to superpose when plotted {\it vs.} $t_{w}/t^{\infty}_{\alpha}$ for all the Laponite concentrations in figure 7(a).\\
\indent We next calculate the non-Gaussian parameter $\alpha_{2}$ (details of the calculations of $\alpha_{2}$ and table T3 of calculated values are supplied in supporting information) associated with the distribution of the $\alpha$-relaxation timescales $\rho_{ww}(\tau)$. In figure 7(b), $\alpha_{2}$ when plotted {\it vs.} $t_{w}/t^{\infty}_{\alpha}$, is seen to superpose for all four Laponite concentrations. It is seen that $\alpha_{2}$ is very small when $t_{w}$ is small. However, $\alpha_{2}$ increases sharply at higher $t_{w}$ for all four Laponite concentrations. In all the superpositions presented here, it is observed that the horizontal shift factor $t^{\infty}_{\alpha}$ decreases rapidly with increasing Laponite concentration (figure 3(b) and the inset of figure 5(a)). The observed superpositions of $\alpha_{1}$ and $\alpha_{2}$, which is achieved without any vertical shift for all the Laponite concentrations, is an additional verification of the self-similarity of the dynamic slowing down process. If $q' =1/t^{\infty}_{\alpha}$ is defined as a rate  at which the system approaches the glass transition, it is seen from the inset of figure 7(b) that $q'$ increases exponentially with concentration. This is connected to our earlier observation that Laponite particles  are trapped in progressively deeper energy wells as the Laponite concentration is increased.
\begin{figure*}[!t]
\includegraphics[width=7in]{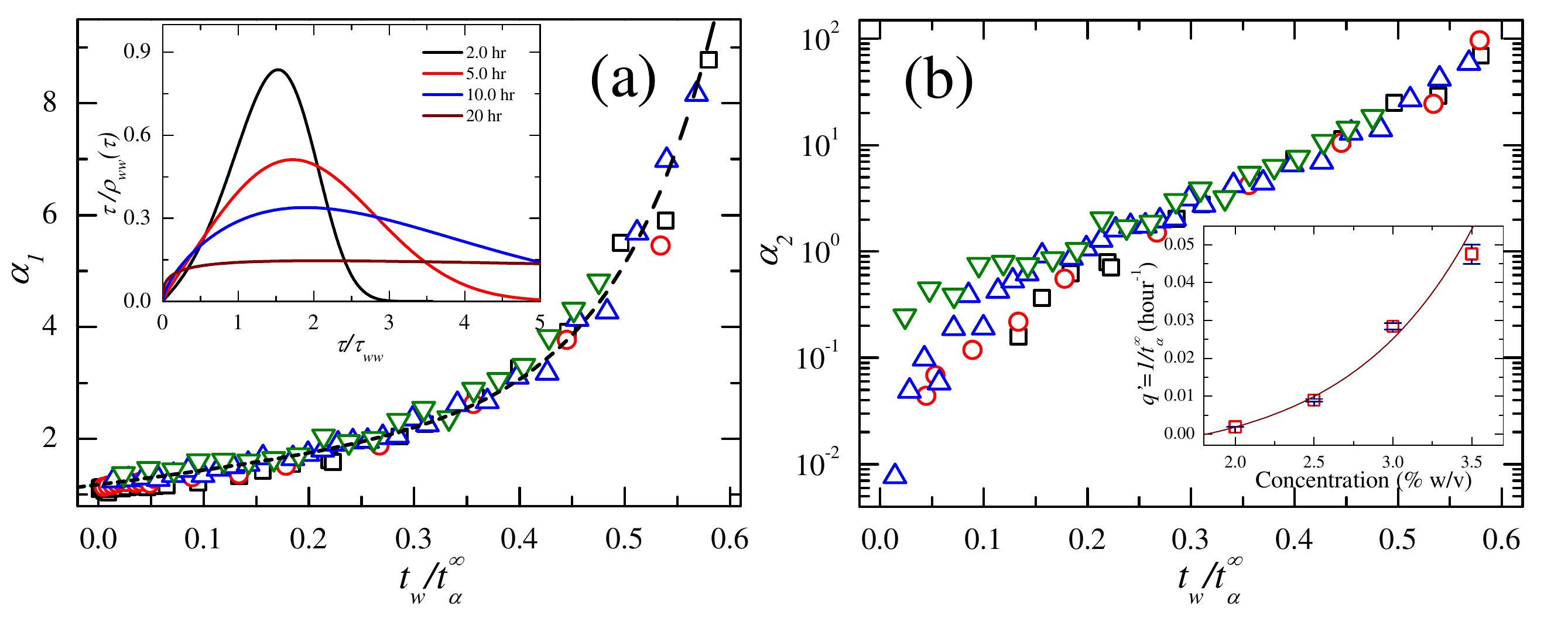}
\caption{(a). Width parameter $\alpha_{1}$ {\it vs.} $t_{w}/t^{\infty}_{\alpha}$ for 2.0\% w/v ($\Box$), 2.5\% w/v ($\circ$), 3.0\% w/v ($\triangle$) and 3.5\% w/v ($\nabla$) Laponite suspensions. In the inset: distributions of the $\alpha$-relaxation timescales plotted for 3.0\% w/v Laponite suspension at 2 hr, 5 hr, 10 hr and 20 hr (from top to bottom). (b) The non-Gaussian parameter $\alpha_{2}$ {\it vs.} $t_{w}/t^{\infty}_{\alpha}$ for 2.0\% w/v ($\Box$), 2.5\% w/v ($\circ$), 3.0\% w/v ($\triangle$) and 3.5\% w/v ($\nabla$) Laponite suspensions. In the inset: the rate $q'$ ($=1/t^{\infty}_{\alpha}$) at which the system approaches the glass transition is plotted {\it vs.} Laponite concentration. The solid line is an exponential fit.}
\label{Figure 7}
\end{figure*}
\section{Conclusions}
In this work, we have extracted the primary and secondary relaxation timescales of aging Laponite suspensions by modeling the intensity autocorrelation functions obtained from dynamic light scattering measurements. We have compared the dynamical slow-down process of these samples with that observed in fragile supercooled liquids. While colloidal suspensions of Laponite approach the glass transition spontaneously with increasing waiting time $t_{w}$, supercooled liquids are obtained by quenching the temperature of a liquid towards its glass transition temperature at a rate that is rapid enough to avoid crystallization. It is proposed in the literature that the faster $\beta$-relaxation process of a supercooled liquid exhibits an Arrhenius temperature-dependence, while the slower $\alpha$-relaxation time exhibits a {\it VFT} temperature-dependence \cite{gotze_mct}. In our work, we have demonstrated remarkably striking similarities in the relaxation processes of soft colloidal suspensions approaching dynamical arrest and fragile supercooled liquids by performing a simple one-to-one mapping between the waiting time since filtration of an aging Laponite suspension and the inverse of the thermodynamic temperature of a supercooled liquid ($t_{w}\leftrightarrow1/T$).\\
\indent We have identified the secondary and the primary relaxation processes of aging Laponite suspensions with, respectively, the $\beta$ and the $\alpha$-relaxation processes of fragile supercooled liquids. We observe here that the secondary relaxation process of aging Laponite suspensions exhibits several signatures of the Johari-Goldstein $\beta$-relaxation process reported in supercooled liquids. Furthermore, we have shown that the evolutions of both the primary and secondary relaxation processes are self-similar with increasing Laponite concentration. Our estimates for the apparent activation energy corresponding to the $\alpha$-relaxation process, the widths of the distributions of the $\alpha$-relaxation timescales and the non-Gaussian parameters characterizing these distributions also confirm the self-similar dynamics of Laponite suspensions with increasing Laponite concentrations.\\
\indent Several simple relations are known to exist among the different temperature scales ({\it eg.} the glass transition temperature $T_{g}$, the Vogel temperature $T_{0}$ and the Kauzmann temperature $T_{k}$) and energy scales ({\it eg.} the activation energy corresponding to the $\beta$ relaxation process $E$) that characterize the glass transition of supercooled liquids.  In this work, we have calculated the glass transition time $t_{g}$ \cite{Angell_Fragility_1991}, and have defined new timescales, such as the timescale corresponding to the secondary relaxation process  $t^{\infty}_{\beta}$, the Vogel time $t^{\infty}_{\alpha}$ and the Kauzmann time $t_{k}$,  to characterize the dynamical slowing down process in Laponite suspensions. We demonstrate  the existence of relations between these timescales that are strongly reminiscent of the relations that were established between the characteristic  temperature/energy scales of supercooled liquids approaching their glass transitions.\\
\indent A comparison of our data with the results obtained for suspensions of hard spheres near the glass transition shows that a suspension of Laponite platelets evolves in the same way with increasing waiting time (equation 3) as a suspension of hard spheres whose volume fraction is increased towards the random close packing fraction of $\phi_{c}$=0.638 \cite{Marshall_Zukoski_JPC_1990}. The slowing down of the dynamics in hard sphere suspensions as $\phi\rightarrow\phi_{c}$ therefore proceeds in the same manner as the slowing down in suspensions of charged Laponite platelets with $t_{w}\rightarrow t^{\infty}_{\alpha}$. The inter-platelet interactions in aging Laponite suspensions evolve spontaneously with waiting time, resulting in an increase in the effective volume fraction and a simultaneous decrease in the accessible volume available to the system. This eventually leads to dynamical arrest. In hard sphere suspensions, the volume fraction plays the same role as the inverse of temperature in the glass transition of molecular glasses and supercooled liquids \cite{Pusey_Van_Megen_Nature_1986,Marshall_Zukoski_JPC_1990}. The mapping ($t_{w}\leftrightarrow1/T$) established here demonstrates that aging Laponite suspensions, hard sphere glasses and fragile supercooled liquids approach their glass transitions in very similar manners. Our study therefore clearly confirms that aqueous suspensions of Laponite are model glass formers.
\section{Additional Material}
\subsection{Supporting Information}
The following figures and tables are supplied in a supporting information file. Figure S1 shows the waiting times associated with the minima in $\tau_{1}$ {\it vs.} Laponite concentration. Figure S2 shows the time-evolutions of the stretching exponent $\beta$ for four Laponite concentrations. A disscussion on the derivations of equations 5 and 6 is included. Figure S3 shows the activation energies associated with the $\beta$- and $\alpha$-relaxation processes for the same concentrations of Laponite. Figure S4 shows the decrease of the stretching exponent $\beta$ with increase in $\log(<\tau_{ww}>/\tau_{1})$. Table T1 tabulates the values of $t^{\infty}_{\alpha}$, $t_{k}$, $t^{\infty}_{\beta}$ and $t_{g}$ for the different Laponite concentrations. Details of the calculations of the width parameter $\alpha_{1}$ and the non-Gaussian parameter $\alpha_{2}$ are discussed. The values of $\alpha_{1}$ and $\alpha_{2}$ with increasing suspension ages for all the Laponite concentrations studied here are supplied in tabular form (tables T2 and T3). Figure S5 shows the intensity autocorrelation function at 60$^{\circ}$ for a 2.5\% w/v sample at four different ages  and fits of this data to equation 1. Figures S6 and S7 show the plots of the fast and the slow relaxation times and fits to equations 2 and 3 respectively at a scattering angle $\theta$ = 60$^{\circ}$. The diffusive behavior of the fast and slow relaxation times in 2.5\% w/v Laponite suspension are shown in figures S8 and S9 respectively.
\section{Acknowledgments}
The authors are grateful to R. Basak for his help with the experiments.
\newpage
\section{}


\section*{Supplementary material}

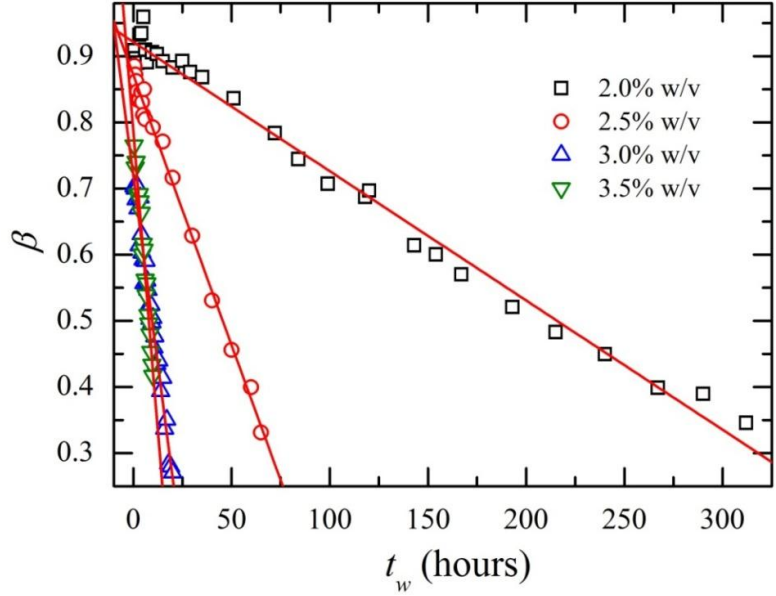

**Figure S2:** The stretching exponent, *β, vs.* waiting time, $t_w$ for 2.0% w/v (□), 2.5% w/v (○), 3.0% w/v (Δ) and 3.5% w/v (∇) Laponite suspensions. The solid lines are linear fits.

<u>Calculations of activation energies</u>

<u>Derivation of equation 5:</u>

For a supercooled liquid, $\tau_1 \sim \exp(E/k_B T)$ [Ngai, *J. Chem. Phys.*, **109**, 6982 (1998)].

Comparison with equation 2 of our paper yields: $\dfrac{E}{k_B T} = \dfrac{t_w}{t_\beta^\infty}$

The one-to-one mapping ($1/T \leftrightarrow t_w$) demonstrated in this paper (equations 2 and 3 of the manuscript, plotted in figures 2(a) and 2(b)) yields the following expression:

$$E = \frac{k_B c_1}{t_\beta^\infty}$$

$E$ can therefore be written in terms of $t_\beta^\infty$. This is plotted in the inset of figure S3. The constant $c_1$ in the above equation has the dimension of [time]×[temperature].

<u>Derivation of equation 6:</u>

According to reference 46:

$$E_{VFT} = k_B \left[\frac{d(\ln \tau)}{d(1/T)}\right]$$

Substituting $\tau = \tau_0 \exp\left(\dfrac{DT_0}{T-T_0}\right)$ [Reference: C. A. Angell, J. Res. Natl. Inst. Stand. Technol. 102, 171 (1997)] in the above expression for $E_{VFT}$, we get, $E_{VFT} = k_B \left[\dfrac{DT_0 T^2}{(T-T_0)^2}\right]$

Substituting $c_2/T = t_w$ and $c_2/T_0 = t_\alpha^\infty$, we obtain

$$E_{VFT} = k_B c_2 \left[ \frac{D}{t_\alpha^\infty \left(1 - t_w / t_\alpha^\infty\right)^2} \right]$$

$E_{VFT}$ can therefore be written in terms of $t_w$ and $t_\alpha^\infty$. This is plotted in the figure S3. The constant $c_2$ in the above equation has the dimension of [time]×[temperature].

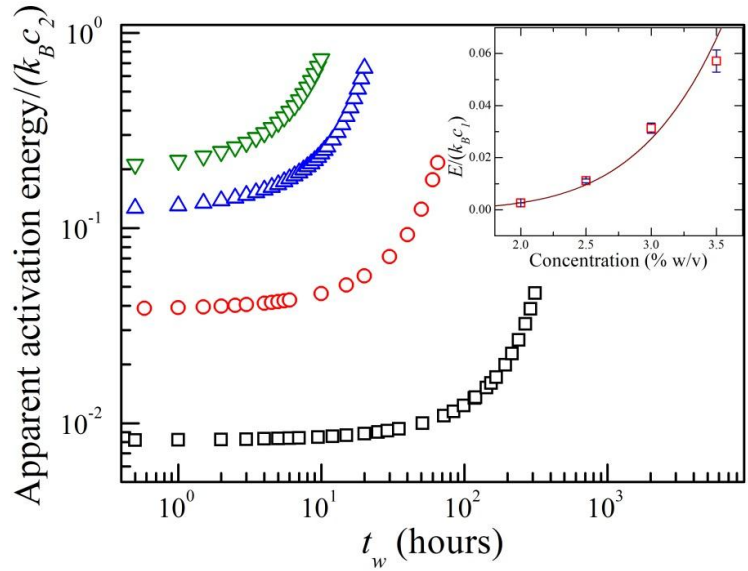

**Figure S3:** The normalized apparent activation energy ($E_{VFT}$) associated with the $\alpha$-relaxation process *vs.* waiting time ($t_w$) for 2.0% w/v (□), 2.5% w/v (○), 3.0% w/v (Δ) and 3.5% w/v (∇) Laponite suspensions. In the inset, the normalized activation energy ($E$) associated with the $\beta$-relaxation process is plotted *vs.* Laponite concentration $c$. The solid line is a power law fit ($E \sim c^{\,5.7 \pm 0.3}$).

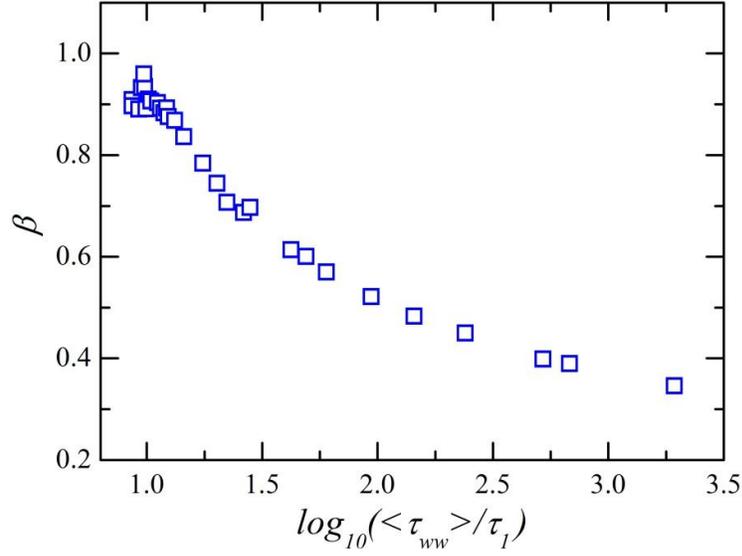

**Figure S4:** The stretching exponent $\beta$ vs. $log_{10}(<\tau_{ww}>/\tau_1)$ for 2.0% w/v Laponite suspension.

**Table T1: Estimates of $t_\alpha^\infty, t_k, t_\beta^\infty$ and $t_g$ vs. Laponite concentration**

| Concentration of Laponite (% w/v) | $t_\alpha^\infty$ (hours) | $t_k$ (hours) | $t_\beta^\infty$ (hours) | $t_g$ (hours) |
|---|---|---|---|---|
| 2.0 | 538.0 | 472.2 | 375.0 | 397.0 |
| 2.5 | 112.3 | 106.9 | 89.8 | 83.1 |
| 3.0 | 35.2 | 35.2 | 31.9 | 25.8 |
| 3.5 | 21.0 | 21.5 | 17.5 | 15.5 |

**Calculation of the width parameter ($a_1$) and non-Gaussian parameter ($a_2$):**

Distribution of $\alpha$-relaxation is given by,

$$\rho_{ww}(\tau) = \frac{\tau_{ww}}{\pi\tau^2} \sum_{k=0}^{\infty} (-1)^k \sin(\pi\beta k)\Gamma(\beta k+1)\left(\frac{\tau}{\tau_{ww}}\right)^{\beta k+1}$$

and the $n^{th}$ moment of the distribution is given by,

$$\left\langle \tau_{ww}^n \right\rangle = \frac{\tau_{ww}^n}{\beta} \frac{\Gamma\left(\frac{n}{\beta}\right)}{\Gamma(n)}$$

The width parameter is given by:

$$\alpha_1 = \frac{\langle \tau_{ww}^2 \rangle - \langle \tau_{ww} \rangle^2}{\langle \tau_{ww} \rangle^2}$$

and the non-Gaussian parameter is given by:

$$\alpha_2 = \frac{3 \langle \tau_{ww}^4 \rangle}{5 \langle \tau_{ww}^2 \rangle^2} - 1$$

**Table T2: Calculations of the width parameter ($\alpha_1$)**

| $t_w$ (hours) 2.0% w/v | $\alpha_1$ 2.0% w/v | $t_w$ (hours) 2.5% w/v | $\alpha_1$ 2.5% w/v | $t_w$ (hours) 3.0% w/v | $\alpha_1$ 3.0% w/v | $t_w$ (hours) 3.5% w/v | $\alpha_1$ 3.5% w/v |
|---|---|---|---|---|---|---|---|
| 0.5 | 1.10462 | 0.58 | 1.13886 | 0.5 | 1.25068 | 0.5 | 1.37439 |
| 1 | 1.12202 | 1 | 1.15897 | 1 | 1.27299 | 1 | 1.46605 |
| 2 | 1.13112 | 1.5 | 1.17539 | 1.5 | 1.29918 | 1.5 | 1.44102 |
| 3 | 1.07323 | 2 | 1.20227 | 2 | 1.27879 | 2 | 1.60044 |
| 4 | 1.07089 | 2.5 | 1.22575 | 2.5 | 1.34573 | 2.5 | 1.61353 |
| 5 | 1.04068 | 3 | 1.22457 | 3 | 1.44056 | 3 | 1.59831 |
| 6 | 1.10304 | 4 | 1.20479 | 3.5 | 1.34722 | 3.5 | 1.64067 |
| 7 | 1.12992 | 4.5 | 1.232 | 4 | 1.4582 | 4 | 1.71128 |
| 9.5 | 1.10856 | 5 | 1.27024 | 4.5 | 1.50602 | 4.5 | 2.04392 |
| 12 | 1.11274 | 5.5 | 1.19652 | 5 | 1.54408 | 5 | 1.94416 |
| 15 | 1.1286 | 6 | 1.28325 | 5.5 | 1.66356 | 5.5 | 1.99879 |
| 20 | 1.14278 | 10 | 1.30975 | 6.5 | 1.64615 | 6 | 2.32306 |
| 25 | 1.12814 | 15 | 1.36033 | 7 | 1.72245 | 6.5 | 2.53988 |
| 29 | 1.1536 | 20 | 1.51513 | 7.5 | 1.80119 | 7 | 2.37135 |
| 35 | 1.16563 | 30 | 1.87861 | 8 | 1.90796 | 7.5 | 2.88049 |
| 51 | 1.22121 | 40 | 2.62192 | 8.5 | 1.95065 | 8 | 3.05582 |
| 72 | 1.32969 | 50 | 3.76481 | 9 | 1.95813 | 8.5 | 3.30927 |
| 84 | 1.43005 | 60 | 5.45427 | 9.5 | 2.02253 | 9 | 3.82171 |
| 99 | 1.54471 | 65 | 10.26433 | 10 | 2.04866 | 9.5 | 4.31219 |
| 118 | 1.61584 | | | 10.5 | 2.35735 | 10 | 4.81554 |
| 120 | 1.5795 | | | 11 | 2.2524 | | |
| 143 | 1.96069 | | | 12 | 2.61647 | | |
| 154 | 2.04236 | | | 13 | 2.66661 | | |
| 167 | 2.26105 | | | 14 | 3.10576 | | |
| 193 | 2.7322 | | | 15 | 3.17541 | | |
| 215 | 3.25971 | | | 16 | 4.13807 | | |
| 240 | 3.91152 | | | 17 | 4.26916 | | |
| 267 | 5.50604 | | | 18 | 5.68258 | | |
| 290 | 5.90128 | | | 19 | 6.97967 | | |
| 312 | 8.77478 | | | 20 | 8.16226 | | |

**Table T3: Calculations of non-Gaussian parameter ($\alpha_2$)**

| $t_w$ (hours) 2.0% w/v | $\alpha_2$ 2.0% w/v | $t_w$ (hours) 2.5% w/v | $\alpha_2$ 2.5% w/v | $t_w$ (hours) 3.0% w/v | $\alpha_2$ 3.0% w/v | $t_w$ (hours) 3.5% w/v | $\alpha_2$ 3.5% w/v |
|---|---|---|---|---|---|---|---|
| 0.5 | -0.24034 | 0.58 | -0.18569 | 0.5 | 0.00769 | 0.5 | 0.24693 |
| 1 | -0.21244 | 1 | -0.15234 | 1 | 0.0488 | 1 | 0.44208 |
| 2 | -0.19752 | 1.5 | -0.12475 | 1.5 | 0.09815 | 1.5 | 0.38702 |
| 3 | -0.28886 | 2 | -0.0785 | 2 | 0.05847 | 2 | 0.75497 |
| 4 | -0.29243 | 2.5 | -0.03697 | 2.5 | 0.18898 | 2.5 | 0.78711 |
| 5 | -0.33765 | 3 | -0.03916 | 3 | 0.38603 | 3 | 0.74938 |
| 6 | -0.24266 | 4 | -0.07439 | 3.5 | 0.19166 | 3.5 | 0.85466 |
| 7 | -0.19965 | 4.5 | -0.02609 | 4 | 0.42451 | 4 | 1.03724 |
| 9.5 | -0.23406 | 5 | 0.04385 | 4.5 | 0.53147 | 4.5 | 2.02054 |
| 12 | -0.22773 | 5.5 | -0.08901 | 5 | 0.61953 | 5 | 1.70427 |
| 15 | -0.2023 | 6 | 0.06826 | 5.5 | 0.91324 | 5.5 | 1.8752 |
| 20 | -0.17934 | 10 | 0.11857 | 6.5 | 0.86864 | 6 | 3.00386 |
| 25 | -0.20344 | 15 | 0.21824 | 7 | 1.06717 | 6.5 | 3.86843 |
| 29 | -0.16157 | 20 | 0.55228 | 7.5 | 1.28284 | 7 | 3.18867 |
| 35 | -0.14165 | 30 | 1.50652 | 8 | 1.59404 | 7.5 | 5.40767 |
| 51 | -0.04568 | 40 | 4.21947 | 8.5 | 1.72437 | 8 | 6.28734 |
| 72 | 0.15727 | 50 | 10.45309 | 9 | 1.74562 | 8.5 | 7.66252 |
| 84 | 0.36368 | 60 | 24.4168 | 9.5 | 1.95025 | 9 | 10.82908 |
| 99 | 0.62136 | 65 | 96.6505 | 10 | 2.03587 | 9.5 | 14.34769 |
| 118 | 0.79297 | | | 10.5 | 3.1348 | 10 | 18.45914 |
| 120 | 0.70384 | | | 11 | 2.74087 | | |
| 143 | 1.7556 | | | 12 | 4.19569 | | |
| 154 | 2.01552 | | | 13 | 4.41496 | | |
| 167 | 2.77215 | | | 14 | 6.54793 | | |
| 193 | 4.71058 | | | 15 | 6.92068 | | |
| 215 | 7.38437 | | | 16 | 13.04328 | | |
| 240 | 11.43881 | | | 17 | 14.01893 | | |
| 267 | 24.86205 | | | 18 | 26.74489 | | |
| 290 | 28.99567 | | | 19 | 42.04245 | | |
| 312 | 69.0407 | | | 20 | 59.04992 | | |

Data acquired at a different scattering angle ($\theta = 60°$)

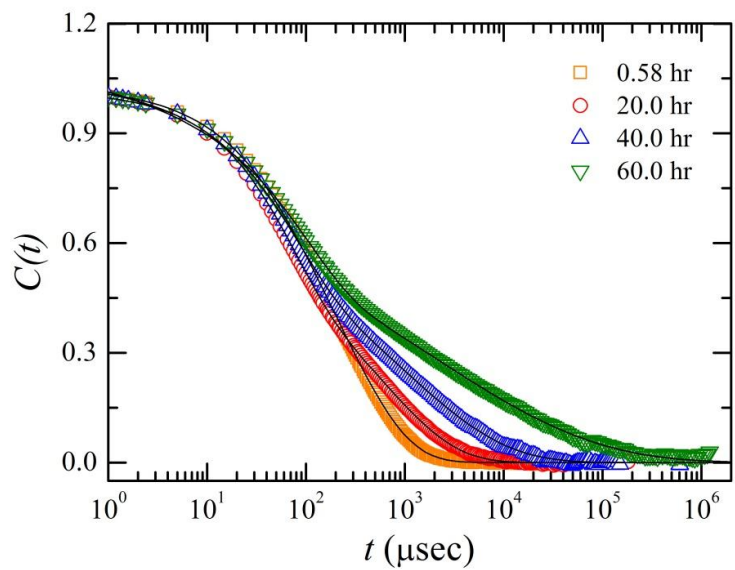

**Figure S5:** Intensity autocorrelation functions for a 2.5% w/v sample at scattering angle $\theta = 60°$ for four different waiting times. The solid lines are fits to equation 1.

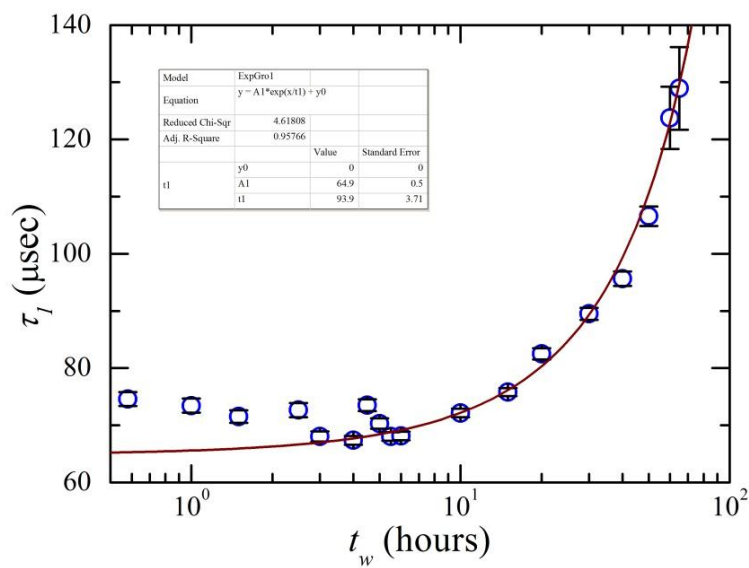

**Figure S6:** The fast relaxation timescales of a 2.5% w/v Laponite sample, extracted from fits of the data plotted in Figure S5 to equation 1, versus time since preparation of the sample. The solid line is a fit to the modified Arrhenius form (equation 2). A decrease in $\tau_1$ at very early times, followed by an eventual increase, as highlighted in Figure 2(a), is also present.

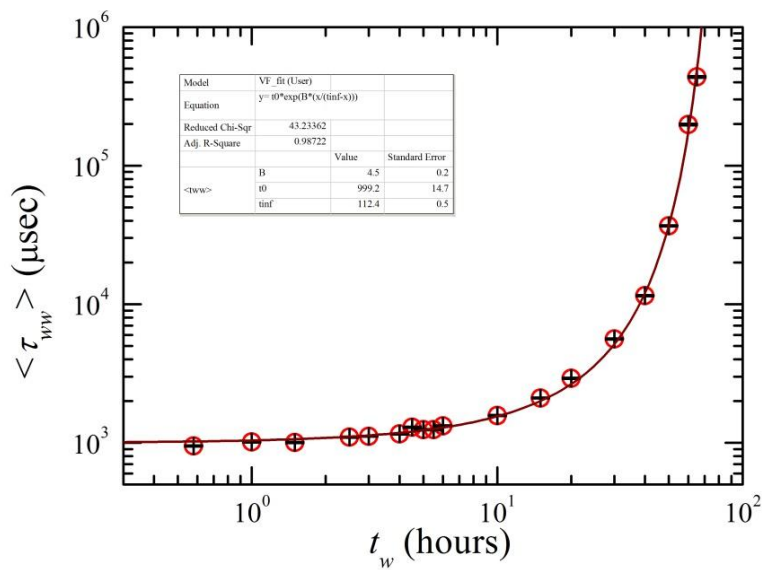

**Figure S7:** The mean slow relaxation timescales of a 2.5% w/v Laponite sample, extracted from fits of the data plotted in Figure S5 to equation 1, versus time since preparation of the sample. The solid line is a fit to the modified VFT form (equation 3).

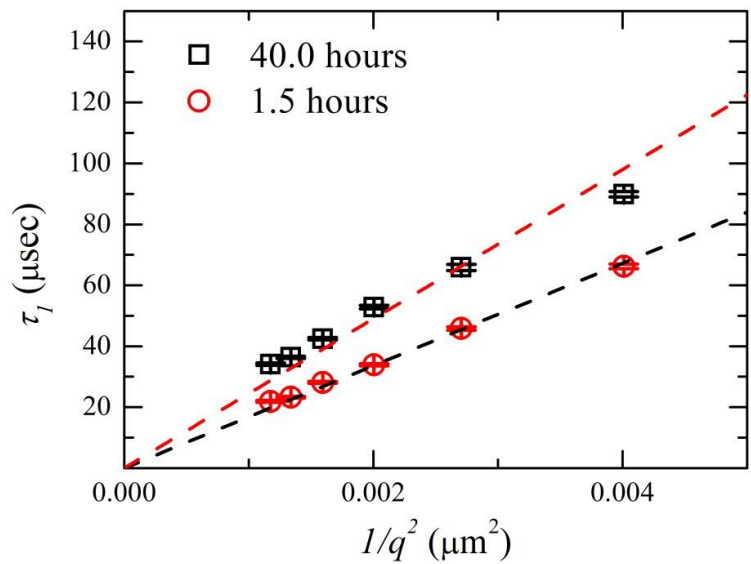

**Figure S8:** The diffusive dynamics of the fast relaxation time ($\tau_1$) is shown above for a 2.5% w/v Laponite sample for two different waiting times $t_w$. The dashed lines are linear fits passing through the origin.

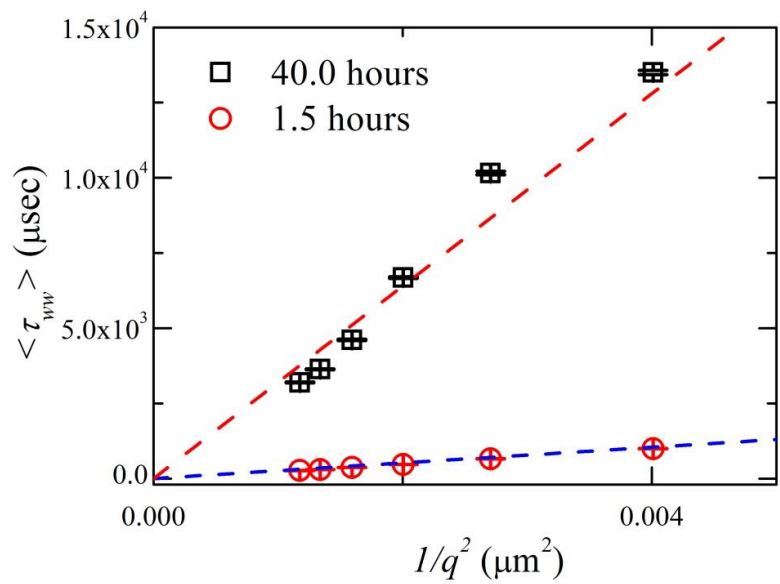

**Figure S9:** The diffusive dynamics of the mean slow relaxation time is shown above for a 2.5% w/v Laponite sample for two different waiting times $t_w$. The dashed lines are linear fits passing through the origin.



\begin{thebibliography}{55}
\bibitem{gotze_mct} W. Gotze and S. J. Sjorgen, {\it Rep. Prog. Phys.}, 1992, {\bf 55}, 241-376.
\bibitem{Ediger_Angell_Nagel_JPC_1996}M. D. Ediger, C. A. Angell and S. R. Nagel, {\it J. Phys. Chem.}, 1996, {\bf 100}, 13200-13212.
\bibitem{Angell_Fragility_1991}C. A. Angell, {\it J. Non-Cryst. Solids}, 1991, {\bf 131-133}, 13-31.
\bibitem{Angell_JP_CS_1988}C. A. Angell, {\it J. Phys. and Chem. of Solids}, 1988, {\bf 49}, 863-871.
\bibitem{Adam_Gibbs_JCP_1965}G. Adam and J. H. Gibbs, {\it J. Chem. Phys.}, 1965, {\bf 43}, 139-146.
\bibitem{Johari_Goldstein_JChemPhys_1970}G. P. Johari and M. Goldstein, {\it J. Chem. Phys.}, 1970, {\bf 53}, 2372-2388.
\bibitem{Johari_Goldstein_JChemPhys_1971}G. P. Johari and M. Goldstein, {\it J. Chem. Phys.}, 1971, {\bf 55}, 4245-4252.
\bibitem{Thayyil_Ngai_Philosophical_Magazine_2008}M. S. Thayyil, S. Capacciolia, D. Prevostoa and K. L. Ngai, {\it Phil. Mag.}, 2008, {\bf 88}, 4007-4013.
\bibitem{Ngai_JCP_1998}K. L. Ngai, {\it J. Chem Phys.}, 1998, {\bf 109}, 6982-6994.
\bibitem{Pusey_Van_Megen_Nature_1986}P. N. Pusey and W. van Megen, {\it Nature}, 1986, {\bf 320}, 340-342.
\bibitem{Marshall_Zukoski_JPC_1990}L. Marshall and C. F. Zukoski, {\it J. Phys. Chem.}, 1990, {\bf 94}, 1164-1171.
\bibitem{bonn_epl}D. Bonn, H. Tanaka, G. Wegdam, H. Kellay and J. Meunier, {\it Europhys. Lett.}, 1999, {\bf 45}, 52-57.
\bibitem{Ruzicka_2004}B. Ruzicka, L. Zulian and G. Ruocco, {\it J. Phys.: Condens. Matter}, 2004, {\bf 16}, S4993-S5002. \\
B. Ruzicka, L. Zulian and G. Ruocco, {\it Phys. Rev. Lett.}, 2004, {\bf 93}, 258301.
\bibitem{ruzicka_review}B. Ruzicka and E. Zaccarelli, {\it Soft Matter}, 2011, {\bf 7}, 1268-1286.
\bibitem{bandyopadhyay_prl}R. Bandyopadhyay, D. Liang, H. Yardimci, D. A. Sessoms, M. A. Borthwick, S. G. J. Mochrie, J. L. Harden and R. L. Leheny, {\it  Phys. Rev. Lett.}, 2004, {\bf 93}, 228302.
\bibitem{schosseler_PRE} F. Schosseler, S. Kaloun, M. Skouri and J. P. Munch, {\it Phys. Rev. E}, 2006, {\bf 73}, 021401.
\bibitem{kaloun_PRE}S. Kaloun, R. Skouri, M. Skouri, J. P. Munch and F. Schosseler, {\it  Phys. Rev. E}, 2005, {\bf 72}, 011403.
\bibitem{tanaka_pre}H. Tanaka, S. Jabbari-Farouji, J. Meunier and D. Bonn, {\it Phys. Rev. E}, 2005, {\bf 71}, 021402.
\bibitem{joshi_lang1}A. Shahin and Y. M. Joshi, {\it Langmuir}, 2012, {\bf 28}, 15674-15686.
\bibitem{Negi_Osuji_PRE_2010}A. S. Negi and C. O. Osuji, {\it Phys. Rev. E}, 2010, {\bf 82}, 031404. 
\bibitem{Angelini_SM_2013} R. Angelini, L. Zulian, A. Fluerasu, A. Madsen, G. Ruocco and B. Ruzicka, {\it Soft Matter}, 2013, {\bf 9}, 10955.
\bibitem{Abou_Bonn_Meunier_PRE_2001}B. Abou, D. Bonn and J. Meunier, {\it Phys. Rev. E}, 2001, {\bf 64}, 021510.
\bibitem{shahin_prl}A. Shahin and Y. M. Joshi, {\it Phys. Rev. Lett.}, 2011, {\bf 106}, 038302.
\bibitem{lcstruik}L. C. E. Struik, {\it Physical aging in amorphous polymers and other materials}, (Elsevier, Houston, 1978).
\bibitem{dhavale_SM}T. P. Dhavale, S. Jadav and Y. M. Joshi, {\it Soft Matter}, 2013, {\bf 9}, 7751-7756.
\bibitem{kovacs_jps}A. J. Kovacs, {\it J. Polym. Sci.}, 1956, {\bf 30}, 131-147.
\bibitem{Strachan_et_al_PRE_2006}D. R. Strachan, G. C. Kalur and S. R. Raghavan, {\it Phys. Rev. E}, 2006, {\bf 73}, 041509.
\bibitem{lacks_prl}D. J. Lacks and M. J. Osborne, {\it Phys. Rev. Lett.}, 2004, {\bf 93}, 255501.
\bibitem{bandy_SM}R. Bandyopadhyay, P. H. Mohan and Y. M. Joshi, {\it Soft Matter}, 2010, {\bf 6}, 1462-1466.
\bibitem{tawari_edge}S. L. Tawari, D. L. Koch and C. Cohen, {\it J.Colloid Interface Sci.}, 2001, {\bf 240}, 54-66.
\bibitem{Ruzicka_PRL_2010}B. Ruzicka, L. Zulian, E. Zaccarelli, R. Angelini, M. Sztucki, A. Moussaid and G. Ruocco, {\it Phys. Rev. Lett.}, 2010, {\bf 104}, 085701.
\bibitem{Ruzicka_Langmuir_2006}B. Ruzicka, L. Zulian and G. Ruocco, {\it Langmuir}, 2006, {\bf 22}, 1106-1111.
\bibitem{Ruzicka_Philosophical_Magazine_2007}B. Ruzicka, L. Zulian and G. Ruocco, {\it Phil. Mag.}, 2007, {\bf 87}, 449-458.
\bibitem{Tanaka_Bonn_PRE_2004}H. Tanaka, J. Meunier and D. Bonn, {\it Phys. Rev. E}, 2004, {\bf 69}, 031404.
\bibitem{Ruzicka_Nature_Materials_2011}B. Ruzicka, E. Zaccarelli, L. Zulian, R. Angelini, M. Sztucki, A. Moussad, T. Narayanan and F. Sciortino, {\it Nat. Mater.}, 2011, {\bf 10}, 56-60.
\bibitem{bandyopadhyay_ssc}R. Bandyopadhyay, D. Liang, J. L. Harden and R. L. Leheny, {\it Solid State Comm.}, 2006, {\bf 139}, 589-598.
\bibitem{Stillinger_Science_1995}F. H. Stillinger, {\it Science}, 1995, {\bf 267}, 1935-1939
\bibitem{Archer_PRL_2011} P. Agarwal, S. Srivastava and L. A. Archer, {\it Phys. Rev. Lett.}, 2011, {\bf 107}, 268302.
\bibitem{Sciortino_et_al_Soft_Matter_2009}F. Sciortino, C. D. Michele, S. Corezzi, J. Russo, E. Zaccarelli and P. Tartaglia, {\it Soft Matter}, 2009, {\bf 5}, 2571-2575.
\bibitem{Kudlik_et_al_EPL_1997}A. Kudlik, C. Tschirwitz, S. Benkhof, T. Blochowicz and E. R$\ddot{o}$ssler, {\it Europhys. Lett.}, 1997, {\bf 40}, 649-654.
\bibitem{Kudlik_et_al_J_Non_Cryst_Solids_1998}A. Kudlik, C. Tschirwitz, T. Blochowicz, S. Benkhof, E. R$\ddot{o}$ssler, {\it J. Non-Crystalline Solids}, 1998, {\bf 235}, 406-411.
\bibitem{Vyazovkin_Dranca_Phermaceutical_Research_2006}S. Vyazovkin and I. Dranca, {\it Pharm. Res.}, 2006, {\bf 23}, 422-428.
\bibitem{Jackle_Rep_Prog_Phys_1986}J. J$\ddot{a}$ckle, {\it Rep. Prog. Phys.}, 1986, {\bf 49}, 171-231.
\bibitem{miyazaki_epl}K. Miyazaki, H. M. Wyss, D. A. Weitz and D. R. Reichman, {\it Europhys. Lett.}, 2006, {\bf 75}, 915-921.
\bibitem{bern_pecora}B. J. Berne and R. Pecora, {\it Dynamic light scattering: With applications to Chemistry, Biology, and Physics}; John Wiley \& Sons: New York, 1975; pp 12-18.
\bibitem{Joshi_JCP_2007}Y. M. Joshi, {\it J. Chem. Phys.}, 2007, {\bf 127}, 081102.
\bibitem{Samim_Ranjini_Langmuir_2013}S. Ali and R. Bandyopadhyay, {\it Langmuir}, 2013, {\bf 29}, 12663-12669.
\bibitem{Lindsey_JCP_1980}C. P. Lindsey and G. D. Patterson, {\it J. Chem. Phys.}, 1980, {\bf 73}, 3348-3357.
\bibitem{larson_rheology}R. G. Larson, {\it The structure and rheology of complex fluids}, Clarendon Press: Oxford, {\bf 1999}.
\bibitem{Sastry_Nature_2001}S. Sastry, {\it Nature}, 2001, {\bf 409}, 164-167.
\bibitem{Introduction_to_Polymer_Viscoelasticity}M. T. Shaw and W. J. MacKnight, {\it  Introduction to polymer viscoelasticity}, Wiley-Interscience, John Wiley \& Sons, 205, pp. 271.
\bibitem{Debenedetti_Nature_2001}P. G. Debenedetti and F. H. Stillinger, {\it Nature}, 2001, {\bf 410}, 259-267.
\bibitem{Goldstein_JCP_1969}M. Goldstein, {\it J. Chem. Phys.}, 1969, {\bf 51}, 3728.
\bibitem{Dixon_Nagel_PRL_1998}P. K. Dixon and S. R. Nagel, {\it Phys. Rev. Lett.}, 1998, {\bf 61}, 341-344.
\bibitem{Bohmer_Ngai_Angell_JCP_1993}R. B$\ddot{o}$hmer, K. L. Ngai, C. A. Angell and D. J. Plazek, {\it J. Chem. Phys.}, 1993, {\bf 99}, 4201-4209.
\end{thebibliography}
\end{document}